# SIMULATION-BASED APPROACH FOR MULTIPROJECT SCHEDULING BASED ON COMPOSITE PRIORITY RULES

Alvarez-Campana, P.; Villafañez, F.; Acebes, F. & Poza, D.

GIR INSISOC - Department of Business Organization and CIM, School of Industrial Engineering, University of Valladolid, Valladolid, 47011, Spain

E-Mail: pablo.alvarezcampana@gmail.com, felixantonio.villafanez@uva.es, fernando.acebes@uva.es, david.poza@uva.es

**Abstract**

This paper presents a simulation approach to enhance the performance of heuristics for multiproject scheduling. Unlike other heuristics available in the literature that use only one priority criterion for resource allocation, this paper proposes a structured way to sequentially apply more than one priority criterion for this purpose. By means of simulation, different feasible schedules are obtained to, therefore, increase the probability of finding the schedule with the shortest duration. The performance of this simulation approach was validated with the MPSPLib library, one of the most prominent libraries for resource-constrained multiproject scheduling. These results highlight the proposed method as a useful option for addressing limited time and resources in portfolio management.

(Received in Month 202_, accepted in Month 202_. This paper was with the authors _ months for _ revisions.)

## 1. INTRODUCTION

Establishing a schedule baseline is a key project planning step because it allows a date to be set for each project activity given all the constraints upon their execution. Traditionally, these constraints have been grouped into two types: time constraints and resource constraints. Time constraints involve considering what other project activities must have been previously executed so that an activity can start. The second constraint ensures that all the resources required for the activity are available in the required quantity and distribution on the date on which the activity is scheduled [1].

Meeting the first constraint is a relatively simple task for which mature methods have been available for several decades (Gantt charts, Critical Path Method – CPM, Program Evaluation and Review Technique – PERT, etc.). These methods, which assume that unlimited resources can be allocated to project activities have been incorporated into most project scheduling software packages [2].

However, in practice, resources are limited in their availability. Obtaining a schedule that also meets the second constraint (i.e. resource constraint) is a far more complex task that involves resorting to complex algorithms which are not necessarily guaranteed to provide an optimum schedule. This scheduling problem, in which project activities must comply with the time and resource constraints, is referred to in the literature as the RCPSP (Resource-Constrained Project Scheduling Problem) [3]. This problem is proven to be NP-hard in the strong sense [4] which means that, in most cases (excluding instances of small size), it is not possible to find the optimal solution (i.e. the optimal project schedule) within an acceptable processing time.

In practice, companies need to manage not only a single project with limited resources, but a set of projects whose execution requires sharing some common resources among the activities from different projects [5, 6]. This problem is known in the literature as the RCMPSP (Resource-Constrained MultiProject Scheduling Problem). The objective of the RCMPSP is to find a portfolio schedule that meets the time and resource constraints, while its total completion time (also known as total makespan, TMS) is minimised [7].

The RCMPSP can be formulated as a mixed integer linear programming model (MILP) [8]. However, similarly to the RCPSP, the RCMPSP is also an NP-hard problem, which implies that there are no known algorithms to obtain an optimal solution to the problem in polynomial time, not even for small-sized problems [9]. For this reason, researchers continue striving to find new approximate solution methods that can effectively solve large-scale scheduling problems (i.e. find a near-optimal solution) quickly [10].

According to a recent survey [11], heuristics based on priority rules are the most representative approximate algorithms to solve the RCMPSP. Priority rules are criteria for determining what activities should receive resources in those cases for which there are not enough resources to run a simultaneous execution due to resource constraints [12]. In this way, the activities that are considered to be of the highest priority receive the amount of resources that they need to be scheduled first. However, if there are not enough resources to execute less prioritised activities, they are scheduled at a later date when the required resources are released by other more prioritised activities [13]. Priority-rule based methods provide sufficiently good solutions in a shorter resolution time in comparison to exact methods [14–17].

Many heuristics based on priority rules provide different schedules when applied to the same project/portfolio (i.e., the same scheduling problem) [18]. The reason is because, when two activities or more meet the criterion associated with the priority rule, the tie-breaker is usually solved randomly. In a chain, this conditions not only the start date of these activities, but also affects the other successive activities until the end of the project. Consequently in this type of heuristics, simulation is usually employed in an attempt to find the schedule with the shortest duration. As the RCMPSP is an NP-hard problem, the solution (i.e., the schedule with the shortest duration) is unknown. However, increasing the number of simulation runs raises the probability of reaching that optimal (but unknown) solution.

In this paper, we present a simulation approach for portfolio scheduling based on a novel way to apply priority rules, which we call Drawers Heuristic. Classic heuristics available in the literature only apply a priority rule, which is usually based on the computation of some numerical parameter, and may be a second tie-breaker criterion if there are activities with the same priority. However, the method herein described goes further and proposes not only a structured way to sequentially apply more than one priority criterion, but also allows the application of more complex priority rules based on the activities' attributes. Depending on whether activities meet one combination of requirements or another, they are firstly classified into different groups (that we refer to as 'drawers'), each with a different priority level. Then a simple priority rule is applied to rank the activities in each drawer. The order of drawers and the order of activities in each drawer determine the order in which activities are attempted to be scheduled. By means of simulation, we obtain different feasible schedules for the portfolio to, therefore, increase the probability of finding the schedule with the shortest duration.

The functionality of the Drawers Heuristic was validated using the MPSPLib library [19] as a benchmark, which is one of the commonest libraries to test resource-constrained multiproject scheduling resolution methods. Despite the heuristic's general purpose (i.e., it needs no *ad hoc* adaptation to be applied to a particular problem), it matches the best-known solution to date with the shortest portfolio duration for 40 % of the MPSPLib problems, and it provides the best-known solution to date for 39 of the 140 problems collected in the library.

The rest of the article is organised as follows. Section 2 describes the implementation of the Drawers Heuristic. Section 3 presents an example of applying the simulation approach. Section 4 shows the results of applying the Drawers Heuristic to the 140 problems in the MPSPLib library. Finally, Section 5 discusses the conclusions drawn from this work.

## 2. METHODS

Priority rule-based heuristics are combined with schedule generation schemes (SGS) for creating feasible schedules that meet both time and resource constraints (i.e., finding solutions to the RCMPSP) [20]. The SGS is the particular procedure used to obtain the list of activities whose predecessors have been completed (i.e., candidate activities) whereas the priority rule is the criterion (or criteria) used to schedule those candidate activities. The main difference between parallel and serial schedule generation schemes (P-SGS and S-SGS, respectively) is the way of iterating while creating the schedule [16]: P-SGS iterates by time-period and S-SGS iterates by activity. P-SGS starts from the first time period and finds all the candidate activities for that time period (and schedules them following the order of the priority rule depending on the availability of resources). In S-SGS, the activity with the highest priority (which is determined by the priority rule) is selected first, and then the earliest time for scheduling is calculated. This process is repeated until all activities are scheduled.

The simulation approach proposed in this paper is based on a Parallel Schedule Generation Scheme (P-SGS). Subsection 2.1 describes the general scheduling process of a P-SGS (which corresponds to the dark-coloured elements in Figure 1). Subsection 2.2 explains the modifications that we made to the general P-SGS approach to obtain better scheduling results (i.e., the elements highlighted in red in Figure 1).

### 2.1 Parallel Schedule Generation Scheme

The project/portfolio time horizon is divided into the minimum indivisible unit of time considered in the project/portfolio (e.g., days, weeks, etc.). We refer to 'scheduling times' as the beginning of these periods. The scheduling process consists of a succession of iterations over successive scheduling times. In each one, the activities that meet certain criteria are scheduled (i.e., the scheduling time is assigned as the definitive start date of those activities).

Following the guidelines of a P-SGS, the process starts with a first iteration, which corresponds to the earliest available scheduling time. Once the first iteration is finished, a new iteration is performed for the next scheduling time, and so on until all the project/portfolio activities are scheduled. During each iteration, the process is subdivided into four steps:

1. Generation of a temporary schedule. This schedule consists of: a set of activities that have already been definitely scheduled in previous iterations (thus their start date is before the current scheduling time); a set of activities that are still pending to be definitely scheduled. The schedule for the latter activities is based on the Critical Path Method (CPM). Consequently, the obtained schedule is temporary because the CPM only considers the activities' precedence relations and ignores their actual resource use

2. Determination of candidate activities. The temporary schedule obtained in the previous step serves as a reference to determine what activities are eligible for definite scheduling at the current scheduling time: those activities whose temporary start date matches the current scheduling time and their predecessors have already been scheduled. We refer to these activities as candidate activities (i.e., unscheduled activities that can start at the time point corresponding to the current iteration)

3. Prioritisation of candidate activities. Here the aim is to determine the order in which candidate activities are attempted to be scheduled. At this point, classic heuristics apply only one priority rule to sort the list of candidate activities. As we explain below, this paper proposes a novel structured way to combine several priority rules by considering some drawers that determine the priority of the candidate activities. The highlighted part of the diagram in Figure 1 (step 3) corresponds to adapting the general scheme to allow the combined priority rules to be applied. Regardless of the method, a prioritised list of candidate activities is obtained at the end of this step.

4. Attempt to schedule candidate activities. Once the prioritised list of candidate activities is obtained, we can attempt to schedule activities one by one following the order of that list. A check is made to see if all the resources required for executing that activity are available throughout its duration. If so, the activity is definitely scheduled and the available resources are updated by deducting the resources used by the activity from the current resources availability. If there are not enough resources, the activity is discarded and, therefore, remains as a candidate activity for the next iteration. The process is repeated with the next activity on the list by considering possible changes in resource availability due to previously scheduled activities. The process continues until all the candidate activities corresponding to the current iteration are processed.

If there are unscheduled activities remain after the current iteration, a time unit is advanced in the scheduling process and a new iteration starts. The process is repeated iteration after iteration until no more unscheduled activities are left.

## 2.2 Drawers Heuristic

What differentiates our approach from other heuristic methods based on priority rules is the way in which the sorted list of candidate activities is obtained (i.e. step 3 in Fig. 1). Traditionally, these candidate activities are ordered according to a certain priority rule (e.g., ascending order of slack or duration or resource use, etc.), and sometimes a second rule is followed to break possible ties. As the selection of different priority rules conditions the performance of a P-SGS heuristic, in this paper we propose a novel structured way to combine several priority rules to increase the likelihood of obtaining a schedule with the shortest possible makespan. The differences between our approach and traditional approaches based on P-SGS are highlighted in red in Figure 1 (see step 3: prioritisation of candidate activities).

Our approach considers a set of several drawers, each with an associated criterion, among which the candidate activities in every iteration are distributed. The criterion to decide whether an activity is inserted into one drawer or another might be based on different attributes. We classify these attributes into three types: (i) the characteristics of the activity itself; (ii) the relation between the activity to the project to which it belongs; (iii) the relation between that activity and the portfolio as a whole.

The activities placed in the first drawer have a higher priority than those classified in successive drawers. Consequently in the third scheduling step (Fig. 1, step 3: prioritisation of candidate activities), each drawer's criterion is checked sequentially per candidate activity following the descending order of priority represented by the order in which drawers are placed (i.e. if an activity does not meet the criterion to be inserted into the first drawer (the drawer with the highest priority), a check is made to see if the activity meets the criterion to be inserted into the second drawer (with a lower priority than the first drawer), and so forth until the activity is inserted in one of the available drawers). Each candidate activity can only belong to one drawer, which is the first one whose criterion is met by the candidate activity. All the activities that are classified in the same drawer have the same priority. The principal part of the priority of activities is determined by the drawer in which it is classified. However, when a drawer contains two activities or more, a second criterion to break any ties between activities must be applied. The simplest way to achieve this is to apply a random ordering of activities to each drawer. The prioritised list of candidate activities is composed of the concatenation of the activities assigned to each drawer by respecting the order of priority of drawers.

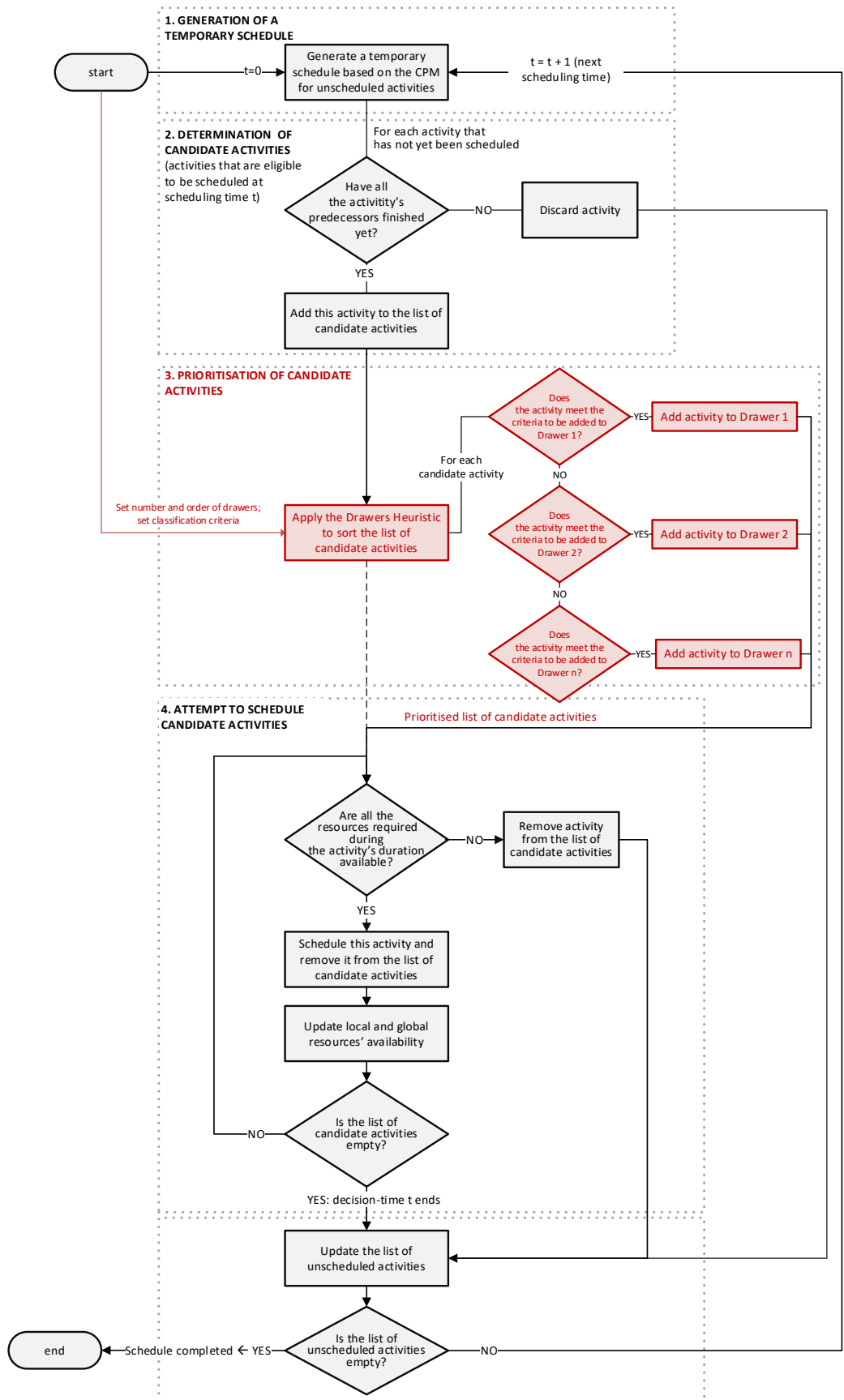

Figure 1: Flow diagram of a P-SGS based on [18, 21].

The randomness introduced by this tie-breaking procedure means that successive runs of the Drawers Heuristic result in different feasible schedules for the same project/portfolio. Consequently to obtain a better schedule for the same problem, we can simulate the scheduling process several times by applying the heuristic repeatedly and maintaining the best solution obtained during the simulation process. Thus the more runs performed, the higher the probability of finding the optimal schedule (remember that being an NP-hard problem, the optimal solution is unknown).

## 3. EXPERIMENTAL WORK: EXAMPLE OF AN APPLICATION

In this section, we present an example of the application of the Drawers Heuristic. As shown in Figure 1, before the iterative scheduling process starts, we need to set the number and order of drawers and the classification criterion associated with each drawer. In this example, we consider a set of four drawers along with some criteria for candidate activities to be included in each drawer:

- First drawer: it contains the candidate activities with total slack of zero and, at the same time, those that belong to the project that ends the last according to the temporary schedule generated in the first step of the scheduling process (Fig. 1). In other words, this drawer includes the candidate activities that currently belong to the critical path of the portfolio
- Second drawer: it contains the activities with a total slack of zero, but do not belong to the project that ends the last according to the temporary schedule. In other words, this drawer contains the activities belonging to the critical path of a project, but do not necessarily belong to the portfolio's critical path
- Third drawer: it accommodates the activities without a total slack of zero, but belong to the project that ends the last
- Fourth drawer: it comprises the activities without zero slack and do not belong to the project that ends the last. In other words, this last drawer comprises all remaining activities that did not fit in the previous drawers

Note that drawers are placed in descending order of priority. Each candidate activity is placed in the first drawer, whose conditions are met by that activity. Once all the candidate activities are classified in their corresponding drawer, some drawers might contain two activities or more. In this case, the order of priority of the activities belonging to the same drawer is randomly determined. As a result, we obtain a list of candidate activities that are sorted according to the priority criteria of the drawer to which they belong, although the order is randomised in each group of activities. In other words, this classification implicitly respects the order in which activities are assigned to drawers (due to the criteria that determine the inclusion of each activity in one drawer or another), but activities in drawers are randomised.

Figure 2 shows an example of assigning candidate activities to the four drawers of this example with an arbitrary iteration during the scheduling process of a 4-project portfolio. The explanation of the illustrative example is as follows. It is assumed that the resource allocation of the previous scheduling times is already processed. This means that the resources required for executing the activities that have already been scheduled are definitely allocated. We go on to describe the resource allocation process for generic scheduling time t. The iteration starts with the generation of a temporary schedule (step 1 in Fig. 2). Please notice that operations in red belong to the critical path based on precedences. Next the candidate activities to be scheduled in the current scheduling time t (i.e. the unscheduled activities that could start at scheduling time t) are determined (step 2).

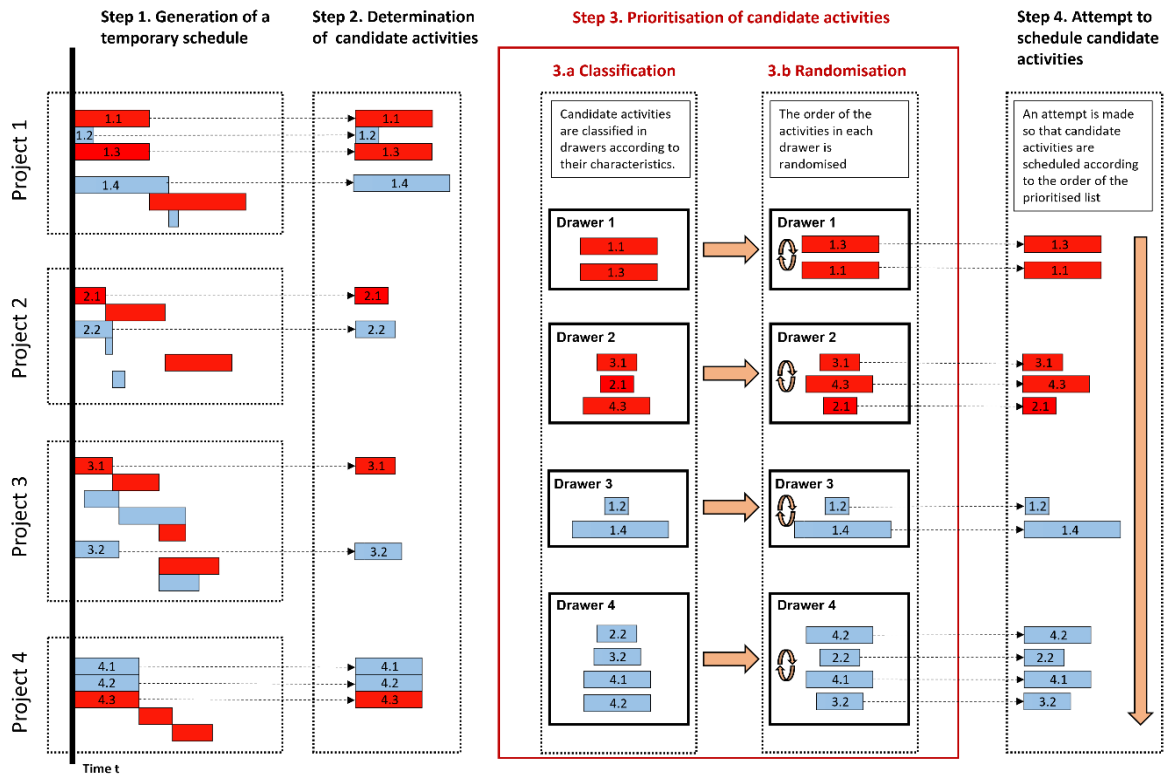

Figure 2: Classification of candidate activities into the four drawers considered in this example.

There are 11 candidate activities in the current scheduling step: Project 1 has four candidate activities (activities 1.1, 1.2, 1.3 and 1.4); Project 2 has two (activities 2.1 and 2.2); Project 3 has two (activities 3.1 and 3.2) and Project 4 has three (activities 4.1, 4.2 and 4.3). Then we obtain a sorted list of candidate activities by means of the Drawers Heuristic (3). The first substep (3.a) consists of classifying the 11 candidate activities into one of the four available drawers. To do so, we take activities one by one and check if the conditions to be included in the first drawer still hold. If not, we check if they can be included in the second drawer, and so forth. According to the temporary schedule, Project 1 is the project that ends the last. Of its four candidate activities, only two (1.1 and 1.3) have zero slack. Consequently, these two activities can be included in the first drawer. The projects that do not end the last (i.e. all the projects except Project 1) have the following critical activities according to the temporary schedule: 2.1, 3.1 and 4.3. These activities can be added to Drawer 2. If we focus again on the project that ends the last (i.e. Project 1), we find it has two activities with non-zero slack (1.2 and 1.4). These activities can be added to Drawer 3. Lastly, we place the remaining candidate activities (2.2, 3.2, 4.1 and 4.2) in Drawer 4. The activities belonging to Drawer 4 have non-zero slack and belong to projects that do not end the last (i.e. all the projects except Project 1). After placing all the candidate activities in a drawer, the activities in each drawer are randomised (substep 3.b, Fig. 2). Finally, an attempt is made to schedule candidate activities according to the prioritised list (step 4).

## 4. SIMULATION AND RESULTS

The fact that the scheduling process of the proposed heuristic presents randomisation (because activities are randomly ordered in drawers) results in different feasible schedules whenever the heuristic is run in the same scheduling problem. Consequently, the use of simulation is essential: a larger number of simulations increases the probability of obtaining a better solution for the same problem.

To validate the performance of the Drawers Heuristic, we resort to the MPSPLib public library [19]. This library contains a set of 140 instances for resource-constrained multiproject scheduling. All these instances are available on the public website http://www.mpsplib.com (last check of address: 18 October, 2023). The problems collected in this library can be used as a benchmark to validate and compare the performance of the new proposed heuristics to the performance of the heuristics put forward by other researchers [22–25]. All 140 problems consist of a portfolio with a variable number of projects (2, 5, 10 or 20) each that have a different number of activities (30, 90 or 120). In addition, the different problems differ from one another by the distinct number of global resources (i.e., the resources shared by all the projects in the portfolio) and local resources (the resources that can be used only by the project to which they belong). MPSPLib also allows researchers to upload the solutions obtained with their heuristics. Thus for all 140 problems in the library, MPSPLib provides a ranking of the best solutions found to date for each problem.

To try to find a portfolio with the lowest total makespan (TMS), 100 runs of the heuristic were performed on all 140 problems in the library. Of the 100 solutions obtained for each problem, that with the shortest TMS was selected as the best solution provided by our heuristic for that problem. The heuristic is configured with the same four drawers, criteria and order as the example shown in Section 3. The algorithm was coded in Microsoft Visual Basic for Applications. All the simulations were implemented on a general-purpose computer with an Intel Core i5 2.4 GHz processor with 6 GB of RAM and Windows 10 Pro as the operating system. The results (i.e., the obtained TMS) for all 140 problems are shown in Appendix 1, where we can find, for the 140 problems, the TMS obtained by the heuristic and that of the method that currently holds the best TMS for that problem. Similarly, the results of applying the heuristic to the 140 problems are uploaded in MPSPLib under the name 'DH/MPR'.

Table 1 summarises the results obtained after running the Drawers Heuristic on all the benchmark instances collected in the MPSPLib library. It shows that this configuration of the heuristic provides the best result to date in 39 of the 140 problems (27 %). It also achieves an equal or better result than the existing ones in 57 of the 140 problems (40 %). It also provides promising results in the problems in which the best result to date is currently achieved by another algorithm. In these cases, the difference between the result obtained by our heuristic and the best solution to date is less than 5 % in 108 of the problems (77 %).

Table I: Performance of the Drawers Heuristic compared to other heuristics in MPSPLib

|  | Number of problems (out of 140) | Percentage |
|---|---|---|
| It offers the best result | 39 | 27 % |
| It is at least as good as the best heuristic | 57 | 40 % |
| The difference with the best result is smaller than 5 % | 108 | 77 % |

In the MPSPLib library, the complexity of all the 140 instances is evaluated based on the indicator AUF (average utilization factor) [26]. This indicator provides, for each resource, the average value of its demand by all the activities in the portfolio in relation to the total available capacity of that resource. When AUF > 1, the resource utilisation is, on average and per time slot, higher than its available capacity. This means that this resource availability constraint will impact the schedule by extending the total portfolio duration. The more resources with a high AUF value, combined with the effect of more projects and more activities per project, the more challenging it will be to obtain the optimal schedule with the shortest duration. It is precisely in these more complex instances where simulation with our heuristic has demonstrated remarkable performance by significantly surpassing the majority of the other proposals available in the library.

# 5. DISCUSSION AND CONCLUSIONS

The literature on portfolio scheduling shows that the development of new methods for solving the RCMPSP is constantly evolving. The main difference between our approach and other heuristics based on a P-SGS is the way to determine the prioritised list of candidate activities (i.e., activities eligible to be scheduled in the current scheduling time) during each iteration of the scheduling process. Unlike other heuristics that use only one priority rule for resource allocation, this paper contributes to the RCMPSP literature in a structured way to sequentially apply more than one priority criterion. This approach, known as Drawers Heuristic, allows more complex priority rules based on activities' attributes to be appreciated.

Nevertheless, it is well-known that priority rules are problem-specific, and there is no universal rule that works best for all projects/portfolios [12, 27, 28]. Despite the significant advancements achieved in recent decades, there remains limited understanding regarding why certain heuristics outperform others [29, 30]. Consequently, managers do not often know which of all the heuristics available in the literature can provide the best result with their specific projects [15]. So the availability of general-purpose heuristics (i.e., heuristics that do not require adaptation to the scheduling problem to which it is applied) is an advantage for managers [18]. Indeed the validation of Drawers Heuristic in scheduling problems with different characteristics (MPSPLib collects benchmark instances with different numbers of activities, projects, resources, etc.) demonstrates that it is an interesting scheduling method for managers. Despite being a general-purpose heuristic, it provides not only the schedules with the shortest duration in 40 % of the MPSPLib problems, but also the best-known solution to date in 39 of the 140 problems collected in the library.

The heuristic herein presented shares some limitations with other heuristics based on priority rules. Increasing the number of simulation runs entails a higher probability of finding a close-to-optimal portfolio schedule. However, more simulations runs also involve longer computational times. The time required to complete 100 runs of the proposed heuristic depended on the complexity of the problem. It ranged from 9.49 seconds (instance with ID 8) to 10.40 hours (instance with ID 120). To prioritise the time required to obtain a feasible schedule over the solution's quality, project managers can resort to commercial software like Microsoft Project or Oracle Primavera. In fact commercial software also employs heuristics based on priority rules for their simplicity [31]. However, the reason why they offer the fastest solution is because they do not exploit the benefits of simulation for obtaining better schedules, but merely offer a good enough solution in which scheduling meets the resource constraints and not necessarily minimising the total makespan. A recent study indicates that the schedules generated using commercial software tend to have longer durations compared to those produced using heuristics, with the difference being particularly significant for complex problems involving many activities and limited resource availability [32]. Yet given the importance of finding a feasible schedule with a shorter duration as a portfolio baseline, the advantages of using simulation outweigh the disadvantages associated with computational burden. The use of simulation allows to find a balance between the quality of the solutions for NP-hard scheduling problems and the required computation time [33, 34].

From an academic point of view, the drawers-based framework used by the heuristic opens the door to a new family of heuristics that use different drawers with distinct priorities to allow new ways to prioritise activities by following the criteria associated with each drawer. In fact, one of the advantages of the proposed heuristic is its scalability: drawers can be added or repositioned to indicate different priority levels for each set of activities. This simulation approach can also be used for reviewing the problem of non-constant resources (by modifying the scheduling constraints), and for also finding the most optimal level of resources in scheduling total time terms (i.e., by solving the dual problem).

## ACKNOWLEDGEMENT

The authors wish to acknowledge to MCIN/AEI, Spanish Government, and to /10.13039/501100011033/FEDER UE, European Union, for the partial support through the PID2022-137948OA-I00 Research Project. This research has been partially financed by the Regional Government of Castile and Leon (Spain) and the European Regional Development Fund (ERDF, FEDER) with grant VA180P20.

APPENDIX

The table below compares the total makespan (TMS) obtained by the heuristic herein proposed (DH/MPR) to the best-known solutions for the 140 RCMPSP instances in the MPSPLib library. When the DH/MPR heuristic holds the best result (39 out of 140 instances), only one row with the DH/MPR heuristic results is included. The cases in which the DH/MPR matches or outperforms the best-known result to date (57 out of 140) are highlighted in grey.

| ID | Method | TMS |
|---|---|---|
| 1 | GA_WS | 187 |
| 1 | DH/MPR | 196 |
| 2 | MATHEU | 108 |
| 2 | DH/MPR | 114 |
| 3 | GA_WS | 242 |
| 3 | DH/MPR | 242 |
| 4 | PSG- | 143 |
| 4 | DH/MPR | 143 |
| 5 | GA_WS | 184 |
| 5 | DH/MPR | 185 |
| 6 | ACO+SM | 61 |
| 6 | DH/MPR | 72 |
| 7 | GA-RK | 58 |
| 7 | DH/MPR | 61 |
| 8 | MAS/PS | 65 |
| 8 | DH/MPR | 65 |
| 9 | CMAS/ES- | 54 |
| 9 | DH/MPR | 56 |
| 10 | CMAS/ES | 58 |
| 10 | DH/MPR | 65 |
| 11 | CMAS | 417 |
| 11 | DH/MPR | 431 |
| 12 | DH/MPR | 275 |
| 13 | DH/MPR | 303 |
| 14 | DH/MPR | 182 |
| 15 | DH/MPR | 407 |
| 16 | GA-RK | 82 |
| 16 | DH/MPR | 85 |
| 17 | GenConst | 78 |
| 17 | DH/MPR | 81 |
| 18 | HYPER | 103 |
| 18 | DH/MPR | 113 |
| 19 | RES | 76 |
| 19 | DH/MPR | 76 |
| 20 | ACO+SM | 86 |
| 20 | DH/MPR | 90 |
| 21 | DH/MPR | 154 |
| 22 | RES | 128 |
| 22 | DH/MPR | 128 |
| 23 | DH/MPR | 209 |
| 24 | RES | 150 |
| 24 | DH/MPR | 150 |
| 25 | GA_WS | 227 |
| 25 | DH/MPR | 241 |
| 26 | RES | 88 |
| 26 | DH/MPR | 88 |
| 27 | MATHEU | 117 |
| 27 | DH/MPR | 127 |
| 28 | RES | 114 |
| 28 | DH/MPR | 114 |
| 29 | RES | 92 |
| 29 | DH/MPR | 92 |
| 30 | CMAS/ES | 121 |
| 30 | DH/MPR | 121 |
| 31 | GT-MAS | 97 |
| 31 | DH/MPR | 99 |
| 32 | GenConst | 163 |
| 32 | DH/MPR | 169 |
| 33 | RES | 122 |
| 33 | DH/MPR | 122 |
| 34 | DH/MPR | 170 |
| 35 | DH/MPR | 224 |
| 36 | CMAS/ES | 79 |
| 36 | DH/MPR | 79 |
| 37 | RES | 114 |
| 37 | DH/MPR | 118 |

| ID | Method | TMS |
|---|---|---|
| 38 | RES | 138 |
| 38 | DH/MPR | 140 |
| 39 | RES | 123 |
| 39 | DH/MPR | 125 |
| 40 | MATHEUR | 151 |
| 40 | DH/MPR | 162 |
| 41 | CMAS/ES | 130 |
| 41 | DH/MPR | 130 |
| 42 | GA_WS | 233 |
| 42 | DH/MPR | 277 |
| 43 | GenConst | 138 |
| 43 | DH/MPR | 141 |
| 44 | DH/MPR | 367 |
| 45 | MATHEUR | 476 |
| 45 | DH/MPR | 479 |
| 46 | HYPER | 155 |
| 46 | DH/MPR | 161 |
| 47 | MATHEUR | 133 |
| 47 | DH/MPR | 143 |
| 48 | MATHEUR | 272 |
| 48 | DH/MPR | 276 |
| 49 | GA_WS | 146 |
| 49 | DH/MPR | 146 |
| 50 | MATHEUR | 108 |
| 50 | DH/MPR | 118 |
| 51 | MATHEUR | 76 |
| 51 | DH/MPR | 87 |
| 52 | GenConst | 202 |
| 52 | DH/MPR | 212 |
| 53 | GA_WS | 230 |
| 53 | DH/MPR | 252 |
| 54 | GA_WS | 201 |
| 54 | DH/MPR | 212 |
| 55 | MATHEUR | 178 |
| 55 | DH/MPR | 191 |
| 56 | MATHEUR | 75 |
| 56 | DH/MPR | 83 |
| 57 | MATHEUR | 164 |
| 57 | DH/MPR | 182 |
| 58 | DH/MPR | 198 |
| 59 | DH/MPR | 184 |
| 60 | DH/MPR | 256 |
| 61 | DH/MPR | 679 |
| 62 | CP-Sym | 170 |
| 62 | DH/MPR | 174 |
| 63 | CMAS/ES | 732 |
| 63 | DH/MPR | 732 |
| 64 | DH/MPR | 260 |
| 65 | MATHEUR | 667 |
| 65 | DH/MPR | 714 |
| 66 | CMAS/ES | 361 |
| 66 | DH/MPR | 369 |
| 67 | DH/MPR | 673 |
| 68 | DH/MPR | 387 |
| 69 | DH/MPR | 271 |
| 70 | MATHEUR | 233 |
| 70 | DH/MPR | 255 |
| 71 | DH/MPR | 436 |

| ID | Method | TMS |
|---|---|---|
| 72 | CP-Sym | 478 |
| 72 | DH/MPR | 481 |
| 73 | CMAS/ES | 127 |
| 73 | DH/MPR | 127 |
| 74 | CP-Sym | 144 |
| 74 | DH/MPR | 148 |
| 75 | MATHEUR | 355 |
| 75 | DH/MPR | 383 |
| 76 | CP-Sym | 120 |
| 76 | DH/MPR | 121 |
| 77 | DH/MPR | 230 |
| 78 | DH/MPR | 267 |
| 79 | DH/MPR | 147 |
| 80 | MATHEUR | 420 |
| 80 | DH/MPR | 469 |
| 81 | CMAS/ES | 187 |
| 81 | DH/MPR | 192 |
| 82 | CP-Sym | 101 |
| 82 | DH/MPR | 108 |
| 83 | MATHEUR | 330 |
| 83 | DH/MPR | 349 |
| 84 | MATHEUR | 159 |
| 84 | DH/MPR | 170 |
| 85 | MATHEUR | 326 |
| 85 | DH/MPR | 361 |
| 86 | CMAS/ES | 72 |
| 86 | DH/MPR | 72 |
| 87 | CMAS/ES | 186 |
| 87 | DH/MPR | 190 |
| 88 | MATHEUR | 328 |
| 88 | DH/MPR | 347 |
| 89 | MATHEUR | 156 |
| 89 | DH/MPR | 168 |
| 90 | MATHEUR | 329 |
| 90 | DH/MPR | 363 |
| 91 | DH/MPR | 561 |
| 92 | CP-Sym | 246 |
| 92 | DH/MPR | 256 |
| 93 | DH/MPR | 622 |
| 94 | CMAS/ES | 317 |
| 94 | DH/MPR | 328 |
| 95 | MATHEUR | 816 |
| 95 | DH/MPR | 897 |
| 96 | RES | 256 |
| 96 | DH/MPR | 261 |
| 97 | DH/MPR | 556 |
| 98 | DH/MPR | 617 |
| 99 | RES | 317 |
| 99 | DH/MPR | 320 |
| 100 | MATHEUR | 808 |
| 100 | DH/MPR | 891 |
| 101 | GA_WS | 748 |
| 101 | DH/MPR | 754 |
| 102 | GA_WS | 169 |
| 102 | DH/MPR | 170 |
| 103 | GA_WS | 370 |
| 103 | DH/MPR | 373 |
| 104 | GA_WS | 461 |
| 104 | DH/MPR | 465 |

| ID | Method | TMS |
|---|---|---|
| 105 | GA_WS | 406 |
| 105 | DH/MPR | 411 |
| 106 | GA_WS | 381 |
| 106 | DH/MPR | 389 |
| 107 | GA_WS | 378 |
| 107 | DH/MPR | 381 |
| 108 | DH/MPR | 134 |
| 109 | DH/MPR | 162 |
| 110 | DH/MPR | 144 |
| 111 | DH/MPR | 368 |
| 112 | DH/MPR | 334 |
| 113 | DH/MPR | 293 |
| 114 | DH/MPR | 964 |
| 115 | DH/MPR | 302 |
| 116 | DH/MPR | 345 |
| 117 | DH/MPR | 850 |
| 118 | DH/MPR | 804 |
| 119 | DH/MPR | 356 |
| 120 | DH/MPR | 289 |
| 121 | MATHEUR | 212 |
| 121 | DH/MPR | 222 |
| 122 | GA_WS | 90 |
| 122 | DH/MPR | 94 |
| 123 | CMAS/ES | 103 |
| 123 | DH/MPR | 106 |
| 124 | MATHEUR | 170 |
| 124 | DH/MPR | 178 |
| 125 | MATHEUR | 105 |
| 125 | DH/MPR | 109 |
| 126 | RES | 95 |
| 126 | DH/MPR | 99 |
| 127 | GA_WS | 211 |
| 127 | DH/MPR | 219 |
| 128 | RES | 101 |
| 128 | DH/MPR | 104 |
| 129 | GA_WS | 169 |
| 129 | DH/MPR | 178 |
| 130 | MATHEUR | 96 |
| 130 | DH/MPR | 101 |
| 131 | CMAS/ES | 581 |
| 131 | DH/MPR | 592 |
| 132 | RES | 614 |
| 132 | DH/MPR | 620 |
| 133 | RES | 292 |
| 133 | DH/MPR | 295 |
| 134 | DH/MPR | 523 |
| 135 | CMAS/ES | 372 |
| 135 | DH/MPR | 377 |
| 136 | RES | 615 |
| 136 | DH/MPR | 624 |
| 137 | CMAS/ES | 576 |
| 137 | DH/MPR | 588 |
| 138 | CMAS/ES | 284 |
| 138 | DH/MPR | 291 |
| 139 | DH/MPR | 519 |
| 140 | CMAS/ES | 355 |
| 140 | DH/MPR | 367 |